\documentclass[12pt]{iopart}

\begin{document}

\title[Continuum calculations for optical modes in icosahedral quasicrystals]{
Continuum elastic sphere vibrations as a model for low-lying
optical modes in icosahedral quasicrystals
}
\author{E. Duval\dag\ , L. Saviot\ddag\ , A. Mermet\dag\ , D. B. Murray\S}
\address{\dag\ Laboratoire de Physicochimie des Mat\'eriaux Luminescents,
Universit\'{e} Lyon I - UMR-CNRS 5620  43, boulevard du 11 Novembre 69622
Villeurbanne Cedex, France}
\address{\ddag\ Laboratoire de Recherche sur la R\'eactivit\'e des Solides,
Universit\'e de Bourgogne - UMR-CNRS 5613, 9 avenue Alain Savary,
BP 47870, 21078 Dijon Cedex, France}
\address{\S\ Department of Physics and Astronomy, Okanagan University College,
3333 University Way, Kelowna, British Columbia, Canada V1V 1V7}
\date{\today}

\ead{lucien.saviot@u-bourgogne.fr}

\begin{abstract}
The nearly dispersionless, so-called ``optical'' vibrational modes
observed by inelastic neutron scattering from icosahedral Al-Pd-Mn
and Zn-Mg-Y quasicrystals are found to correspond well to modes of a
continuum elastic sphere that has the same diameter as the corresponding
icosahedral basic units of the quasicrystal. When the sphere is
considered as free, most of the experimentally found modes can be
accounted for, in both systems. Taking into account the mechanical
connection between the clusters and the remainder of the quasicrystal
allows a complete assignment of all optical modes in the case of
Al-Pd-Mn. This approach provides support to the relevance of clusters in
the vibrational properties of quasicrystals.
\end{abstract}

\submitto{\JPCM}


\section{Introduction}
Quasicrystals are long-range ordered materials whose diffraction
patterns present symmetries incompatible with translational
invariance \cite{Jan94,Boissieu94}. Most structural analyses
of icosahedral phases have shown that they can be described as
a quasiperiodic packing of group of atoms or ``clusters'' with
a local icosahedral symmetry (\textit{e.g.} pseudo-Mackay
icosahedra).

Experimental electron density maps of crystal approximants have
shown that more electrons are localized on clusters than
between clusters \cite{Kirihara}, supporting
this idea of cluster building blocks. From the vibrational point
of view, the experimental evidence
for the resulting expected phonon
confinement is less clear. Inelastic 
neutron scattering (as well
as inelastic X-ray scattering \cite{Krisch02}) on several
icosahedral phases revealed that the excitation spectrum can be 
split into two regimes: acoustic
and optical \cite{Boissieu93,Bou95,Shi02}. The acoustic regime is
characterised by well defined longitudinal and transverse modes which
display a linear dispersion in the low wavevector $q$ range. The high
energy part of the excitation spectrum displays somewhat broadened 
bands ($\sim$4 meV width) of dispersionless excitations or optical 
modes. The cross-over between the two regimes is abrupt and occurs 
for wavevectors $q$ between 3.5 and 6~nm$^{-1}$, \textit{i.e.} for a 
wavelength of the order of $D_{cl}$ ($D_{cl}$: cluster diameter). This 
shows up as a rapid broadening of the acoustic 
modes and an intermixing of phonon modes in the different branches. Although
no gap opening has been observed, the energy position of the optical
bands matches the crossing of the acoustic branch with pseudo-zone
boundaries as defined by Niizeki \cite{Niizeki90}. These observations
are in general agreement with the theoretical calculations of Hafner and
Kraj\v{c}i for icosahedral quasicrystals  \cite{Haf93} although these
calculations do not provide a clear description of the optical modes.

The low lying energy range of the so-called optical
modes, which features acoustic-optical crossing points in the dispersion
curves, suggests some substantial interaction with the acoustic
modes. Along these lines, a hybridization scheme between acoustic and
optical modes has recently been proposed \cite{Kat04}. In this latter
work, the so-called optical
modes correspond to modes which are confined in the nanometric clusters.

Following the idea that the clusterlike nature of quasicrystals is
crucial for both sound waves \cite{Kat04} and electronic properties, we
propose a detailed identification of the so-called optical
modes in quasicrystals, based on the approach of fundamental vibrational
modes of nanospheres. 
The first kind of analysis that we apply 
(\textit{free} sphere model, section \ref{free})
is to consider clusters as the elemental structural
entities of quasicrystals.  In this way, we can
account for the low lying optical modes, since the
model gives frequencies which compare well with
those experimentally found both for 
\textit{i}-Al-Pd-Mn and \textit{i}-Zn-Mg-Y.
These modes are spheroidal and/or
torsional vibrations of the corresponding clusters. Although crude,
this description turns out to already provide satisfactory 
estimates
of the optical mode frequencies in a model without any freely
adjustable parameters. In a second stage (sphere weakly
coupled to a matrix, section \ref{coupling}), the cluster picture is
given further realistic input by considering the vibrational coupling
with a surrounding matrix, to tentatively account for cluster/cluster
interactions and/or clusters overlapping. This latter model is found to
match more completely with the experimental 
observations, particularly
with respect to the damping parameters of the considered modes.

\section{Vibrations of nanospheres and scattering of acoustic waves}

For almost twenty years, studies have been carried out
on the vibrational modes of approximately spherical
nanometric clusters or nanocrystals embedded in glasses
or in macroscopic crystals using Raman or Brillouin light
scattering \cite{Duv86,Mari88,Fer95,Sav98,Pal99,Por01}. For these
embedded nanocrystal systems, the observed low-lying modes correspond
well to those of a free continuous-medium nanosphere of density
$\rho_{cl}$, longitudinal speed of sound $v_{cl}^L$, transverse speed
of sound $v_{cl}^T$ and diameter $D_{cl}$. The vibrational modes of
such a sphere were studied for the first time by Lamb \cite{Lam82}. The
modes of a free sphere are characterized by a polarization index $p$
that denotes either spheroidal (SPH) or 
torsional (TOR) modes, the usual
quantum numbers $\ell$ and $m$ of spherical harmonics and the harmonic
index $n$. Among the spheroidal modes the simplest are the spherical
($\ell=0$), dipolar ($\ell=1$) and quadrupolar ($\ell=2$) modes. The
mode frequency $\nu_{p \ell m n}$ is inversely proportional to the
diameter $D_{cl}$ and proportional to the (longitudinal or transverse)
sound speed $v_p$ in the material of the nanosphere:

\begin{equation}
 \label{qua1}
 \nu_{p \ell m n}=\frac{S_{p \ell n} v_p}{D_{cl}}
\end{equation}

 \noindent
where $v_{TOR} = v_{cl}^{T}$, $v_{SPH} = v_{cl}^{L}$ and $S_{p \ell n}$ is a
constant for modes of type $(p,\ell,n)$. The numerical values of $S_{p
\ell n}$ were found long ago \cite{Lam82} for a free vibrating sphere.

In a recent work \cite{Mur04,LSDM04}, it was shown that a sphere
embedded inside a matrix scatters acoustic waves more efficiently when
their frequencies match certain values which turn out to be generally
very close to the free sphere eigenfrequencies. More precisely, it is
possible to calculate the position and width of these resonances using
the so-called Complex Frequency Model (CFM). The interesting point
is that these resonances do not depend on the nature of the incident
acoustic waves (plane waves, spherical waves, longitudinal, transverse,
\ldots ). Therefore, we can expect these resonances to be important
even when many spheres are involved. This is similar to the
``hard'' and ``soft'' scatterers picture invoked to interpret sound mode
broadening in quasicrystals \cite{Kat04}.

In the following, we will apply this continuum mechanical model
to very small clusters (less than 2~nm diameter). Also the application
of continuum type boundary conditions at the interface between the
icosahedral cluster and the rest of the quasicrystal is an idealization.
A more accurate treatment would need to consider an atomic level
description of the cluster and its interface \cite{Gav03}. However,
despite these limitations, our numerical results are in good agreement
with experiments.

In order to illustrate the applicability of nanosphere mode analysis to
quasicrystals, we first consider the results of the free sphere model
(because it requires fewer parameters) for Al-Pd-Mn and Zn-Mg-Y. Then,
in order to account for a more realistic situation and a more complete
vibrational pattern, we examine the case where Al-Pd-Mn clusters are
weakly coupled to a surrounding matrix.

\section{\label{free}Free sphere model}
\subsection{Al-Pd-Mn}

The shape of the icosahedral clusters in Al-Pd-Mn is nearly-spherical.
The number of atoms per cluster ($\simeq 51$) can be considered as
sufficient to approximate the lowest-energy confined modes by those
of a continuum sphere. Using the approach described in  \cite{Mur04},
the frequencies of the vibrational modes were calculated for a
continuous-medium free nanosphere with a diameter equal to the
size of the cluster, i.e. $D_{cl}=1.0$~nm \cite{Jan96,Bou92,Yam03}.
Both longitudinal and transverse sound speeds in the clusters were
approximated with the bulk sound speeds in the Al-Pd-Mn quasicrystal,
i.e. $v_{cl}^{L}$ = 6500~m/s and $v_{cl}^T$ = 3500~m/s \cite{Ama92}.
Table~\ref{Al-Pd-MnFree} displays the so-obtained frequency values for
the fundamental ($n=0$) and the first harmonic ($n=1$) of each type of
mode (SPH or TOR) and angular momentum $\ell$.

\begin{table}
\caption{\label{Al-Pd-MnFree}Calculated vibrational
energies for a free cluster of
 Al-Pd-Mn of diameter $D_{cl}$ = 1.0 nm. Corresponding experimentally
 observed optical modes are indicated in superscripts.}
\begin{indented}
\item[]\begin{tabular}{@{}llll}
\br
 & &Spheroidal&Torsional\\
 & &E (meV)&E (meV)\\
\mr
$\ell=0$&$n=0$&$22.8^{O4}$&  \\
&$n=1$&52.2&  \\

$\ell=1$&$n=0$&$16.3^{O3}$& 26.6 \\
&$n=1$&32.7& 42.0 \\

$\ell=2$&$n=0$&$12.2^{O2}$&$ 11.5^{O2}$ \\
&$n=1$&$23.1^{O4}$& 32.9 \\
\br
\end{tabular}
\end{indented}
\end{table}

The optical modes of Al-Pd-Mn quasicrystals observed by inelastic neutron
scattering  \cite{Bou95} have the following approximate energies:
$E_{O1}\simeq 7$ meV, $E_{O2}\simeq 12$ meV, $E_{O3}\simeq 16$ meV
and $E_{O4}\simeq 24$ meV, with hardly any dependence on wavevector $q$.
The results of the free sphere model calculations
presented in table~\ref{Al-Pd-MnFree} 
permit the following
assignments: 
\begin{enumerate}
 \item[$O2$:] the $E_{O2}\simeq 12$ meV mode corresponds to the
 spheroidal and torsional $\ell=2$ fundamental modes. Both
 of these modes are five-fold degenerate.
 \item[$O3$:] the $E_{O3}\simeq 16$ meV mode corresponds to the
 triply degenerate dipolar $\ell=1$ spheroidal mode.
 \item[$O4$:] the $E_{O4}\simeq 24$ meV mode corresponds to the
 spherical $\ell=0$ mode and also to the five-fold degenerate
 quadrupolar $\ell=2,n=1$ spheroidal mode.
\end{enumerate}

It turns out that the \textit{free} sphere calculations are able
to provide an accurate assignment for all of the optical mode
energies except for the lowest one ($O1$) (as described in section
\ref{coupling}, calculations taking into account the cluster-matrix
interaction allows the ``missing'' mode $O1$ to be accounted for).
For each of these modes, we might expect the inelastically
scattered intensities to scale with the corresponding total number
of modes, taking into account their degeneracy. Such
tentatively appears to be the case when examining the
reported lineshapes \cite{Bou95}. However, to make a detailed
comparison would require a quantitative analysis of the inelastic
scattering intensities of individual modes, which is beyond the
scope of this work.

In order to further assess the relevance of the \textit{free} sphere
description to the low energy optical modes in quasicrystalline
structures, we now examine the case of Zn-Mg-Y.

\subsection{Zn-Mg-Y}

The icosahedral quasicrystal Zn-Mg-Y has clusters of
diameter $D_{cl}$ = 1.2 nm \cite{Shi02}.
We calculated the energies of a free spherical cluster having the
same diameter ($D_{cl}$ = 1.2 nm) as the Zn-Mg-Y clusters 
and the same sound speeds as bulk Zn-Mg-Y : $v_{cl}^L$ = 4800 and 
$v_{cl}^T$ = 3100~m/s.  \cite{Shi02}
The results are given in table~\ref{Zn-Mg-Y}.

\begin{table}
 \caption{\label{Zn-Mg-Y}Calculation of vibrational energies
 for a free sphere of Zn-Mg-Y of diameter $D_{cl}$  = 1.2~nm.}
\begin{indented}
\lineup
\item[]\begin{tabular}{@{}llll}
\br
 & &Spheroidal&Torsional\\
 & &E (meV)&E (meV)\\
\mr
$\ell=0$&$n=0$&$12.4^{O2}$&  \\
&$n=1$&31.6&  \\

$\ell=1$&$n=0$&$10.8^{O2}$&$ 19.6^{O3}$ \\
&$n=1$&20.7& 31.0 \\

$\ell=2$&$n=0$&$\08.9^{O1}$&$\08.5^{O1}$ \\
&$n=1$&$15.5^{O3}$& 24.3 \\

\br
\end{tabular}
\end{indented}
\end{table}

Once again, the calculated vibrational energies are found to compare
well with those of the optical modes measured through inelastic
neutron scattering, \textit{i.e.} $E_{O1}\simeq 8$ meV, $E_{O2}\simeq
12$ meV and $E_{O3}\simeq 17$ meV  \cite{Shi02} :
\begin{enumerate}
 \item[$O1$:] the $E_{O1}\simeq 8$ meV mode corresponds to the
   five-fold degenerate spheroidal and torsional $\ell=2$ modes (whose
   frequencies are usually almost the same),
 \item[$O2$:] the $E_{O2}\simeq 12$ meV mode corresponds to the
 spherical $\ell=0$ and the spheroidal dipolar $\ell=1$ modes,
 \item[$O3$:] the $E_{O3}\simeq 17$ meV mode corresponds to the
 five-fold degenerate quadrupolar $\ell=2,\, n=1$ and the triply
 degenerate torsional $\ell=1$ modes.
\end{enumerate}

Much like the results obtained for Al-Pd-Mn, the
\textit{free} sphere model is found to provide a sound
description of all the optical modes in Zn-Mg-Y. The above
assignments are restricted to the low energy range
($E<30$ meV) where modes were unambiguously identified
experimentally.

It is worth noting that from group theoretical selection
rules \cite{Duv92}, only the spherical and spheroidal quadrupolar modes
can be observed by Raman scattering, as experiments confirm (these
selection rules can be broken for anisotropic materials or under
resonant excitation). No such selection rules exist for inelastic
neutron scattering. Besides, it should be noted that the degeneracies of
the aforementioned modes (in any case for $\ell \leq 2$) are not lifted
by lowering the symmetry from spherical to icosahedral. This provides
a supplementary justification for using spheres to approximate the
icosahedral clusters.

The present calculation predicts undamped vibrational modes whereas the
observed excitations are somewhat broadened ($\sim$4~meV). This may
be at least partly attributed to a variation of actual quasicrystal
cluster masses, since structural studies showed that there are several
chemical decorations on the same cluster skeleton or by introducing
a cluster matrix interaction. Moreover, as evidenced in the Al-Pd-Mn
case with the $O1$ mode, the free sphere model does not account for
all vibrational modes when the clusters are embedded in a matrix, even
if the cluster/matrix coupling is weak. In the following section, it
is shown that calculations taking into account the cluster/matrix
interaction can account for the ``missing'' $O1$ mode and for the finite
width of the modes.

\section{\label{coupling}Sphere weakly coupled to a matrix: the Al-Pd-Mn case}

The mode frequencies of embedded
clusters can be calculated knowing the elastic properties of both the clusters 
($\rho_{cl}$, $v_{cl}^L$ and $v_{cl}^T$)
and the neighboring matrix
($\rho_{m}$, $v_{m}^L$ and $v_{m}^T$) and by assuming the
continuity
of the stresses and of the
atomic displacements upon crossing the cluster/matrix 
interface \cite{Mur04}. From Murray and Saviot \cite{Mur04},
it is possible to determine the spectral width of a confined mode as 
functions of the elastic constants and densities of both the 
cluster and the matrix. As expected, the weaker the contrast of elastic 
constants and densities between cluster and matrix, the stronger the 
delocalization of the cluster mode in the matrix and the larger its 
spectral width. Furthermore, from the same authors \cite{Mur04},
taking into account the cluster/matrix interaction leads to ``extra
modes'' that involve the coupled motions of the cluster and the
surrounding matrix. These modes will be referred to as mixed
cluster-matrix modes.

Similarly to the \textit{free} sphere modes, the frequencies of the
cluster-matrix modes are inversely proportional to the cluster diameter.
The more notable modes of this type are the \textit{librational}
torsional (TOR,$\ell=1$) mode and the \textit{rattling} spheroidal
dipolar (SPH,$\ell=1$) mode. Both of these modes correspond to
semi-rigid oscillations of the cluster linked to its surroundings.
The frequencies of these last two kinds of modes are zero for a free
cluster.

To \textit{phenomenologically} model the weakness (\textit{i.e.}
mechanical flexibility) of the joint between the nanosphere representing
the nanometric clusters and the matrix, we introduce an additional
``X-layer'' consisting of a softer (in terms of elastic constants) and
lighter medium, as in a previous work \cite{PortalesPRB02}. Note at this
point that such a simplified
model only aims to account for a possible
partial delocalization of the cluster vibrational wavefunction, due to
its interaction with an environment. The calculation of the
clusters eigenmodes requires the knowledge of the X-layer parameters
(density $\rho_X$, longitudinal speed of sound $v_{X}^L$, transverse
speed of sound $v_{X}^T$ and thickness $d_X$). Since these parameters
are \textit{a priori} unknown, they were numerically adjusted so that
both librational and rattling mode energies fit with that of the lowest
energy optical mode ($E_{O1}\simeq 7$ meV).

In table~\ref{Al-Pd-Mn} the calculated X-layer model vibrational
energies and the phonon full widths at half maximum (FWHM) are given
using the same cluster as before with an intermediate X-layer. The
parameters of the X-layer giving rise to a lowest optical mode energy of
7 meV are sound speeds $v_{X}^L$ = 2000 ~ m/s and $v_{X}^T$ = 1000~m/s,
mass density $\rho_X$ = 3.4~g/cm$^3$ and thickness $d_X$ = 0.05~nm.
Although the X-layer may be viewed as the so-called
``glue''  regions
interconnecting the clusters \cite{Jan94,Boissieu94,Kirihara}, the
so-obtained X-layer parameter values are only meant to model a weak
linking between the cluster and its surroundings; they cannot be given
a firm physical meaning.
The surrounding matrix was modelled as a thick shell surrounding the
X-layer with an infinite radius. The sound speeds and mass density of
the surrounding matrix were set equal to those of the cluster, hence of
the bulk quasicrystal. Note that the so-called ``matrix modes''
\cite{Mur04} are not given here because of their huge damping; they
are irrelevant to the present investigation.

\begin{table}
 \caption{\label{Al-Pd-Mn}Calculated energies for a Al-Pd-Mn cluster, diameter
 $D_{cl}$ = 1.0~nm, weakly bonded to the rest of the quasicrystal by a soft X-layer.}
\begin{indented}
\lineup
\item[]\begin{tabular}{@{}llllll}
\br
 & &\multicolumn{2}{c}{Spheroidal}&\multicolumn{2}{c}{Torsional}\\
 & &E (meV)&FWHM (meV)&E (meV)&FWHM (meV)\\
\mr
$\ell=0$&$n=0$&$24.5^{O4}$& 1.4 & & \\
&$n=1$&51.4& 0.8 & & \\

$\ell=1$&$n=0$&$\06.6^{O1}$& 2.4 & $\07.0^{O1}$&0.6 \\
&$n=1$&$16.9^{O3}$& 0.4 & $26.2^{O4}$&0.4 \\
 &$n=2$&32.1& 0.6& & \\

$\ell=2$&$n=0$&$14.6^{O2}$& 1.6 & $13.2^{O2}$&0.6 \\
&$n=1$&$22.8^{O4}$& 0.6 & 31.9&0.6 \\

\br
\end{tabular}
\end{indented}
\end{table}

The comparison of the free cluster vibrational energies
(table~\ref{Al-Pd-MnFree}) with those of the cluster linked to the
surroundings (table~\ref{Al-Pd-Mn}) shows that the cluster/matrix
interaction does not significantly change the vibrational frequencies
(the parameters of the X-layer and of the matrix have only a weak
effect on most of the optical mode energy positions). The main
differences originate from non-zero FWHM's and the appearance of
cluster-matrix mixed modes (rattling and librational ones) in the
latter case. These mixed modes are relatively strongly localised on
the cluster. Thanks to the X-layer model, the lowest energy optical
mode $O1$ can now be assigned to the triply degenerate librational
mode (TOR,$\ell=1$,$n=0$) and also to the triply degenerate spheroidal
rattling mode. The assignment of the other modes, $O2$ and $O4$,
remains unchanged, while the $O3$ mode now corresponds to the first
harmonic ($n=1$) of the triply degenerate dipolar $\ell=1$ spheroidal
mode. Obviously, the lowest-energy mode depends relatively strongly
on the cluster-surroundings bonding, the properties of which are not
well known. Therefore, at the present stage, it should be considered
cautiously.

The approach we have proposed in this article
relies on no further assumption than a cluster based structure that has
widely been reported for both \textit{i}-AlPdMn and \textit{i}-ZnMgY.
Cluster modes are a \textit{natural} consequence of a structure having a
nanometric relief. In that sense, equivalent modes are expected to occur
for non-iscosahedral cluster structures (for instance cylinders exhibit
radial breathing modes like those observed in carbon nanotubes). So far,
most inelastic neutron studies and computer simulation studies have
assigned the low energy optical modes to the dynamics of single atoms.
Such assignment derives from the ability of both techniques to probe the
dynamics inherent to a particular species, yet they do not 
provide information about
the precise nature of the modes which involve the considered atom. 
The cluster origin of the optical modes is not necessarily in contradiction
with single atom dynamics: the participating ratio of a single atom to
a cluster mode is expected to depend on the nature of the atom and its
location within the cluster as well as on the type of cluster mode.

\section{Conclusion}

The optical modes observed by inelastic neutron scattering in
icosahedral Al-Pd-Mn and Zn-Mg-Y quasicrystals correspond well to the
modes of free continuous-medium nanospheres having the same diameter
as their building clusters and the same sound speeds as in the bulk
quasicrystal.

The model of the nanosphere is expected to be valid only for the
lowest-energy modes corresponding to the fundamental modes or, at most
to their first harmonics, \textit{i.e.} modes for which the separation
between maxima and minima of the vibrational wavefunction is much larger
than the interatomic distance.

The linking of the clusters with their surroundings (as
phenomenologically modelled by the X-layer model) partially
delocalizes the confined modes in the matrix. Due to the
interaction, through the ``glue'' regions, between neighbouring
clusters, cluster modes of same symmetry are expected to couple,
thereby establishing a coherence among them.

\ack The authors are very grateful to M. de Boissieu, R. Currat, S. Francoual
and E. Kats for illuminating discussions.

\end{document}